\documentclass[aps,prx,twocolumn,showpacs,psfig,superscriptaddress,longbibliography]{revtex4-1}

\usepackage{textcomp}
\usepackage{times}
\usepackage{graphicx}
\usepackage{float}
\usepackage{latexsym,amsmath,amssymb,bm,euscript}
\usepackage{color}
\usepackage{subfigure}
\usepackage{epstopdf}
\usepackage[colorlinks=true,linkcolor=blue,citecolor=blue]{hyperref}
\usepackage{hyperref}
\usepackage{soul}
\usepackage[normalem]{ulem}
\usepackage{mathrsfs}
\usepackage{amsmath}
\usepackage{lettrine}
\usepackage{xspace}
\usepackage{textcomp}

\begin{document}

\title{Deconfined Quantum Critical Point: A Review of Progress}

\author{Yi Cui}
\affiliation{School of Physics, Renmin University of China, Beijing, 100872, China}

\author{Rong Yu}
\affiliation{School of Physics, Renmin University of China, Beijing, 100872, China}

\author{Weiqiang Yu}
\email{wqyu\_phy@ruc.edu.cn}
\affiliation{School of Physics, Renmin University of China, Beijing, 100872, China}

\begin{abstract}

Deconfined quantum critical points (DQCPs)
have been proposed as a class of continuous quantum phase transitions
occurring between two ordered phases with distinct symmetry-breaking patterns,
beyond the conventional framework of Landau-Ginzburg-Wilson (LGW) theory.
At the DQCP, the system exhibits emergent gauge fields, fractionalized
excitations, and enhanced symmetries.
Here we review recent theoretical and experimental progress on exploring
DQCPs in condensed matter systems. We first introduce
theoretical advancements in the study of DQCPs over the past twenty years,
particularly in magnetic models on square lattices, honeycomb lattices, kagome
lattices,
and one-dimensional spin chains.
We then discuss recent progress on experimental realization of DQCP in
quantum magnetic systems.
Experimentally, the Shastry-Sutherland model, realized in SrCu$_2$(BO$_3$)$_2$, offers a particularly promising platform for realizing DQCPs. The magnetic frustration inherent to this model drives phase transitions between two distinct symmetry-breaking states: a valence bond solid (VBS) phase and a N\'{e}el antiferromagnetic phase. Remarkably, SrCu$_2$(BO$_3$)$_2$ has provided the first experimental evidence of a proximate DQCP through a field-induced Bose-Einstein condensation, transitioning from the VBS state to the N\'{e}el state.
Nevertheless, the direct experimental realization of a DQCP remains a significant challenge. Despite this, it offers a promising platform for exploring emergent phenomena through quantum phase transition in low-dimensional quantum systems.


\end{abstract}

\maketitle

\section{Introduction}

\subsection{Order-disorder quantum phase transition}

Quantum phase transition (QPT) and quantum criticality are foundational topics
in condensed matter physics~\cite{Sachdev_QPT_2011,Sachdev_2023}.
A cornerstone framework in describing phase transitions
is the Landau-Ginzburg-Wilson (LGW) theory. According to
the LGW theory, phase transitions are generally
characterized by symmetry breaking, where an order parameter is introduced to
distinguish between different phases.
Across a phase transition, the symmetry of the system may
be spontaneously broken, causing the order parameter to
vary from zero to a non-zero value. The transition could be either
continuous (of second-order type) or discontinuous (of first-order type).
The LGW theory has effectively explained numerous
types of phase transitions, such as the ferromagnetic-to-paramagnetic
transitions, as well as phase transitions without symmetry breaking such as liquid-gas transitions.

When a phase transition occurs at zero temperature, usually termed as a QPT,
the driving force is not thermal fluctuations but quantum fluctuations.
If this QPT is second-order, a QCP emerges.
At the QCP, the correlation length diverges, signifying that the entire system becomes critically correlated,
and this divergence underpins the unique scaling behavior associated with
the universality class,
which is dictated, according to the LGW theory, by the symmetry group and
the spatial dimension of the system.

Within the LGW paradigm, QCPs are typically associated with transitions between an ordered phase and a disordered phase.
For example, by tuning external parameters such as field, pressure, or doping, a system may transition from an antiferromagnetic (AFM)
to a paramagnetic (PM) phase, which represents one of the most common types of QCPs.
However, when a continuous transition occurs between two distinct ordered
phases, such as two different types of ordering, the LGW theory
imposes a fundamental constraint: the unbroken symmetry group of one phase must
be a subgroup of the other. If this condition is violated, the transition is
expected to be of first-order type, as the continuous evolution of an order parameter is
prohibited within the LGW framework.

While LGW theory has been instrumental in explaining many types of phase transitions, its limitations have become
evident in specific systems, particularly in the case of the topological phases, and
also the deconfined quantum critical point (DQCP) described below. These phenomena require alternative theoretical frameworks.

\subsection{Unconventional continuous order-order quantum phase transition}

In 2004, Senthil {\it et al.}  proposed that a continuous second-order transition may appear between two ordered states with different types of symmetry breaking,
under the influence of Berry phase effects, leading to a phenomenon known as a DQCP, which indeed exceeds the Landau framework~\cite{Senthil_Science_2004}.
At the DQCP, the system exhibits enhanced symmetries beyond those specified by
the
microscopic Hamiltonian, along with emergent deconfined fractionalized
excitations and gauge fields~\cite{Senthil_PRB_2004,Senthil_PRB_2006,
Nahum_PRL_2015,Shao_Science_2016,Wang_PRX_2017,Qin_PRX_2017,SuL_PRB_2024}.
While the phases on
both sides of the transition remain conventional and ``confined'', the
critical point is characterized by deconfined degrees of
freedom~\cite{PoilblancD_PRB_2006,Ma_PRB_2018,Senthil_review_2024}.
For example, in magnetic systems, magnons with integer spins dominate the excitations in the ordered phases
on
both sides of the transition, whereas at the critical point, the
excitations turn out to be fractionalized as free spinons with half-integer
spins.

\begin{figure}[t]
\centering
\includegraphics[width=7cm]{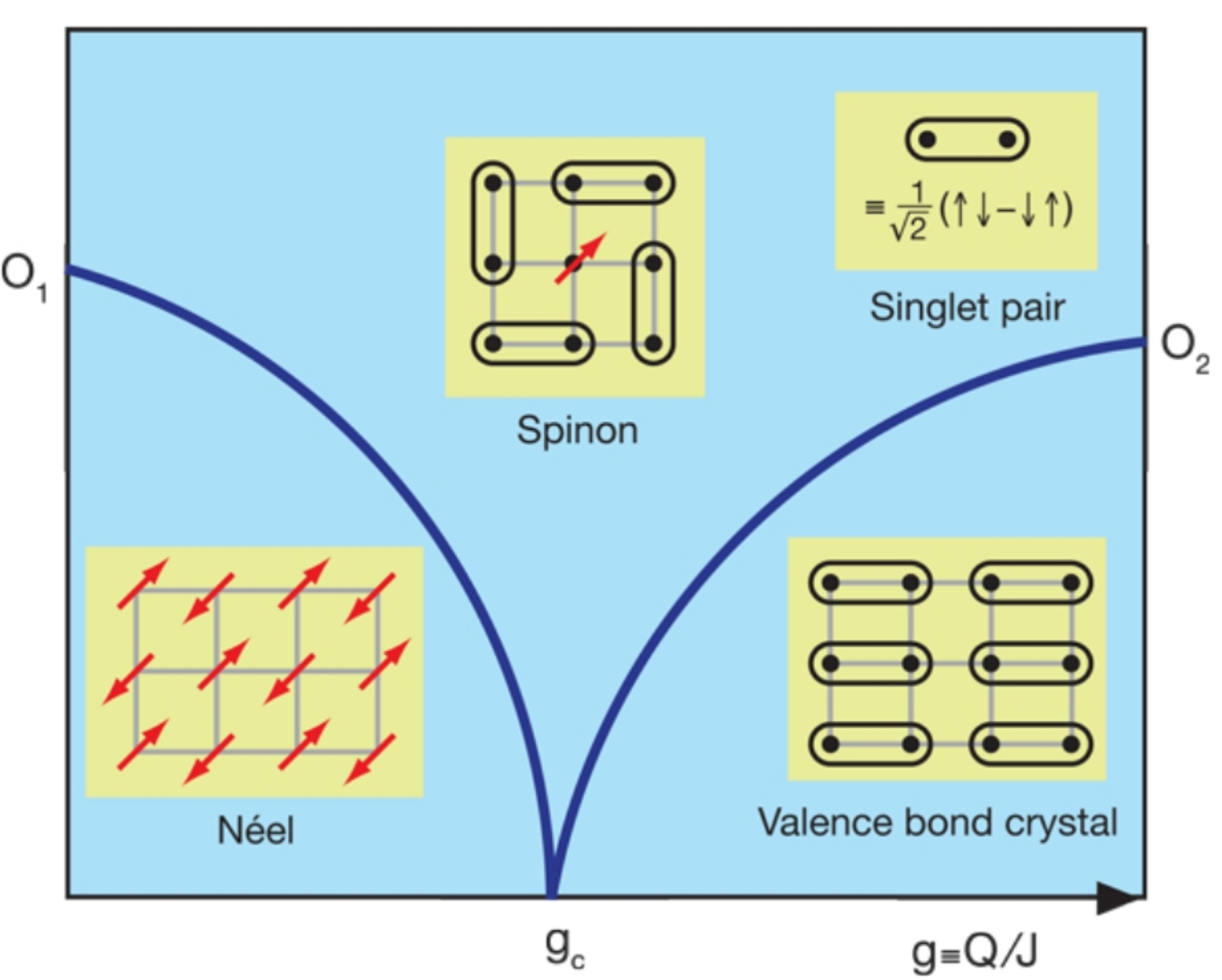}
\caption{\label{dqcppd} Schematic phase diagram across a DQCP between two ordered ground states.
At the DQCP, the N\'{e}el order parameter O$_1$ and the valence bond crystal order parameter O$_2$ on either side
vanish continuously, and deconfined spinons, shown in the inset as free spin-1/2 objects in a sea of
singlet pairs, emerge. Adapted from Ref.~\cite{Singh_physics_2010}.
}
\end{figure}

The DQCP was initially proposed to be realized in a square lattice magnetic system with easy-plane anisotropy.
Figure~\ref{dqcppd} illustrates the phase diagram, in which a DQCP separates the N{\'e}el AFM phase and the valence bond solid/crystal (VBS) phase.
Each phase exhibits distinct symmetry-breaking behaviors: the AFM phase spontaneously breaks a continuous spin-rotational symmetry,
while the VBS phase, where spins pair up to form local singlet bonds (as illustrated in Fig.~\ref{dqcppd}), spontaneously breaks
the translational symmetry with a shorter bond on the singlet bond.
The lowest excitations in both phases are magnons, or specifically, spin waves in the N{\'e}el phase and triplet excitations in the VBS phase.
Remarkably, these two types of order parameters meet at a single transition point, which is a continuous phase
transition~\cite{Nahum_PRL_2015,Mcclarty_NP_2017}. On
the two sides of the transition, the symmetries are O(2) and Z(4),
respectively. By tuning a parameter to the critical point, an emergent O(4) symmetry arises, accompanied by deconfined fractional spinon excitations.
Hereafter, the DQCP framework has been extended to describe continuous phase transitions between other ordered states with different spontaneously broken symmetries.

\section{Theoretical approach to DQCP}

In the last 20 years, DQCPs have attracted significant attention, particularly in exploring the nature of phase transitions, quantum criticality~\cite{Shao_Science_2016,Yang_PRB_2022}, duality phenomena~\cite{Janssen_PRB_2017,Wang_PRX_2017,Mengzy_PRX_2017,Qin_PRX_2017,You_PRB_2021,ShuYR_PRL_2022}, emergent symmetries~\cite{Nahum_PRL_2015,Serna_PRB_2019,Sun_CPB_2021,Xi_JPhysA_2022,Guzc_PRL_2024}, and quantum spin liquid (QSL)~\cite{LiuWY_PRX_2022,Yang_PRB_2022,wang_CPL_2022,Federico_arxiv_2023,wangfa_PRB_2024,LiuWY_Scibull_2024}.

DQCPs have been extensively explored in low-dimensional systems, including models on two-dimensional (2D)
square~\cite{Sandvik_PRL_2007,KaulRK_PRL_2012,QiY_PRB_2014,WangL_PRB_2016,FerrariF_PRB_2018,Lee_PRX_2019,ShackletonH_PRB_2021},
honeycomb~\cite{Pujari_PRL_2013,LiuZH_PRL_2022}, triangular~\cite{ Pimenov_PRB_2017,Jian_PRB_2018},
and Kagome lattices~\cite{Zhang_PRL_2018,ZhangXF_PRB_2024},
as well as one-dimensional (1D) spin chains~\cite{Huang_PRB_2019,HuangRZ_PPP_2020,YangS_PRE_2021,LiCX_PRB_2023,LiuZH_PRL_2023}.
Due to inherent competing interactions, these models are predicted to host
intriguing quantum phases and phase transitions, although experimental verification remains a significant challenge.

\subsection{$J-Q$ model on the square lattice }

DQCP remained a theoretical concept until the introduction of the square lattice $J$-$Q$ model~\cite{Sandvik_PRL_2007}, which established a practical framework for exploring this phenomenon. The Hamiltonian of the $J$-$Q$ model is given by:
\begin{equation}
{\mathcal H} = J\sum_{\langle ij \rangle}\mathbf{S}_i \cdot \mathbf{S}_j - Q
\sum_{\langle ijkl \rangle}(\mathbf{S}_i \cdot \mathbf{S}_j -1/4)(\mathbf{S}_k \cdot \mathbf{S}_l -1/4),
\label{jq}
\end{equation}
where $\langle ij \rangle$ represents nearest-neighbor sites, $\langle ijkl \rangle$ denotes the corner sites of each square in
the lattice, and
$\mathbf{S}_i$ is an $S = 1/2$ spin operator at site $i$.
The parameter $J$ ($>$$0$)  corresponds to the nearest-neighbor AFM Heisenberg coupling, while $Q$ ($>$$0$) represents a four-spin interaction.
In this model, the $J$ term favors the N\'{e}el order, whereas the competing
$Q$ term promotes the formation of a VBS state with local plaquette singlets~\cite{Sandvik_PRL_2007}.

The evolution of the ground states can be investigated through quantum Monte Carlo (QMC) simulations.
Accordingly, the transition between the  N{\'e}el AFM order and a columnar dimer order appears to be continuous ~\cite{Sandvik_PRL_2007}.
The properties of this hypothetical critical point have been further explored through field-theoretical approaches~\cite{Nahum_PRX_2015},
suggesting that this transition might, in fact, represent a peculiar, weakly first-order phase transition. Nevertheless, the scaling exponent $\eta$
exhibits anomalies, indicating that this critical point may be a multicritical point~\cite{Zhao_PRL_2020}.

\subsection{Continuum excitations at DQCP}

\begin{figure}[t]
\centering
\includegraphics[width=8.5cm]{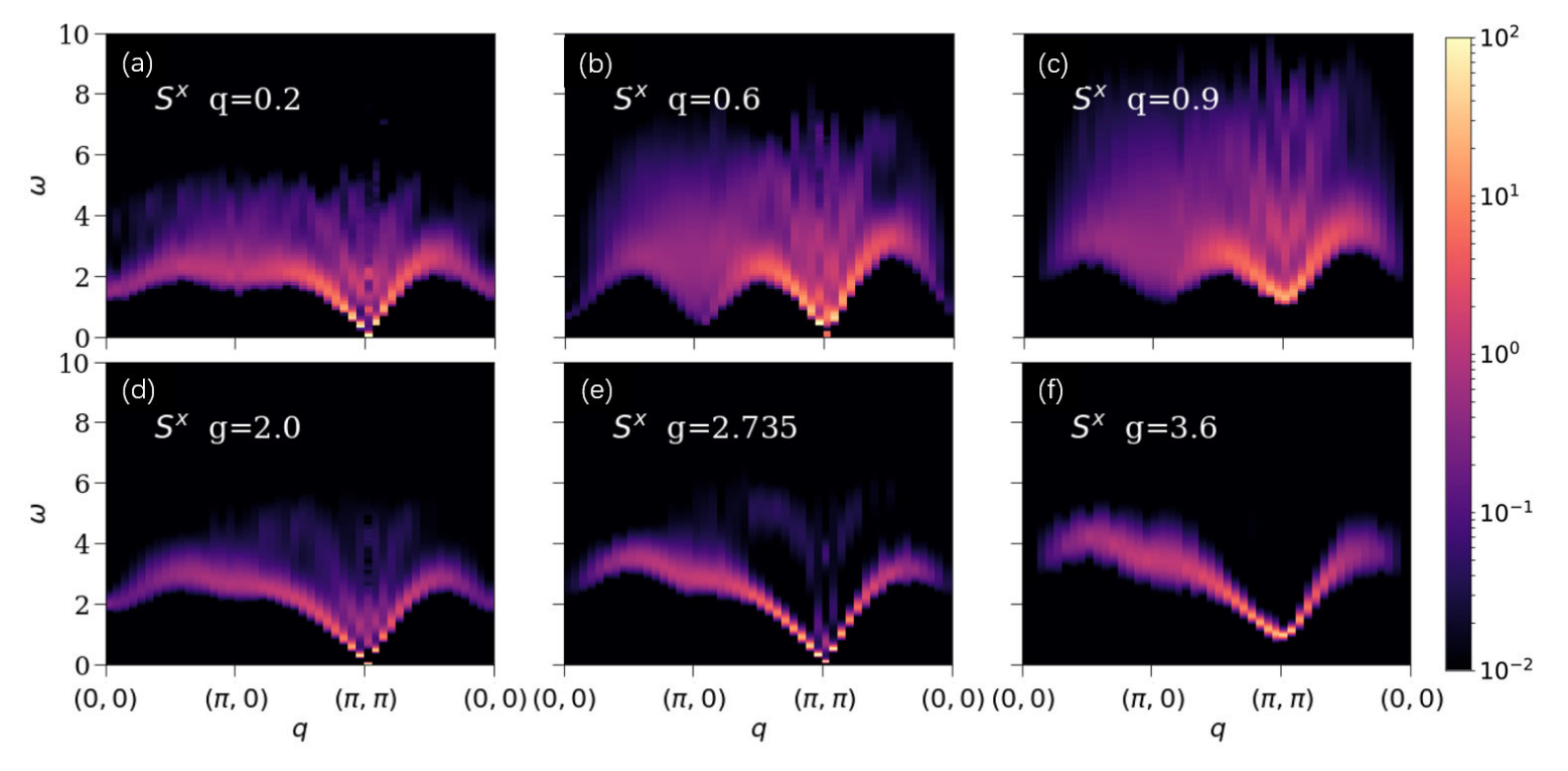}
\caption{\label{dynamic}Dynamic spin structure factors $S^x(\bm{q}, \omega)$ obtained from QMC-SAC calculations for the EPJQ model (a-c)
and the EP-J$_1$J$_2$ model (d-f) with different parameter values. Here (a) and (d) are inside the AFXY phase; (b) and (e) are close to
the DQCP and 3DXY transition point, respectively; and (c) and (f) are inside the VBS phase. Adapted from Ref.~\cite{Ma_PRB_2018}.
}
\end{figure}

If DQCP exists in condensed matter materials, it is expected to exhibit distinct features that
set them apart from conventional QPTs. In Reference~\cite{Ma_PRB_2018}, the spin excitation spectrum of the easy-plane J-Q (EPJQ) model, proposed to host a DQCP, was calculated. For comparison, a conventional QPT was simulated by artificially introducing symmetry breaking in the lattice. The results are illustrated in Fig.~\ref{dynamic}.

Figures~\ref{dynamic}(a)-(c) show the dynamic spin structure factors $S^x(\bm{q}, \omega)$ obtained from QMC simulations combined with stochastic analytic continuation (SAC)  for the EPJQ model
across different phases by tuning the value of $q$ ($q=Q/J$), with a DQCP proposed at $q=0.6$.
In contrast, Figs.~\ref{dynamic}(d)-(f) depict the corresponding progression in the EP-J$_1$J$_2$ model,
where $J_1$ and $J_2$ represent the nearest and the next-nearest neighboring interactions,
which represents a conventional QPT.
Here (a) and (d) are within the AFM phase, (c) and (f) are inside the VBS phase,
and (b) and (e) are close to the DQCP and 3DXY transition point, respectively.

In Fig.~\ref{dynamic}(a), the system resides in the AFM phase, where the spin-wave spectrum displays a gapless Goldstone mode at the $(\pi,\pi)$ point, with spectral broadening arising from spin-wave interactions. As the system approaches the DQCP [Fig.~\ref{dynamic}(b)], a distinct continuum emerges in the $(\bm{q}, \omega)$ space across the entire energy spectrum.
This continuum, arising from fractionalized spinon excitations, is a hallmark of DQCP. Importantly, the spectral broadening at this stage is significantly greater than what would be expected from the critical fluctuations of a conventional QPT, as illustrated in Fig.~\ref{dynamic}(e).

Another critical signature lies in the intensity distribution along the lower edge of the energy spectrum, particularly from $(\pi, 0)$ to $(\pi, \pi)$. This variation reflects the fractionalization of spin waves into spinons, which are not free particles but are strongly coupled to an emergent gauge field. This coupling leads to a shift in spectral intensity and highlights the interplay between the matter field (spinons) and the emergent gauge field. This phenomenon mirrors the confinement-deconfinement transitions familiar from high-energy physics, providing a remarkable bridge between condensed matter and particle physics. These results not only underscore the experimental signatures of DQCP, but also demonstrate how high-energy physics concepts can manifest in quantum materials.

\subsection{Shastry-Sutherland model on the square lattice}

\begin{figure}
\centering
\includegraphics[width=4cm]{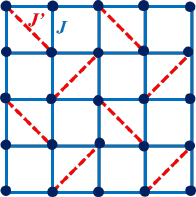}
\caption{\label{ssm} The orthogonal dimer lattice of the Shastry-Sutherland model,
featuring intradimer coupling $J'$ (dashed lines) and nearest-neighbor interdimer coupling $J$ (solid lines)~\cite{Shastry_1981}.
}
\end{figure}

\begin{figure}
\centering
\includegraphics[width=8.5cm]{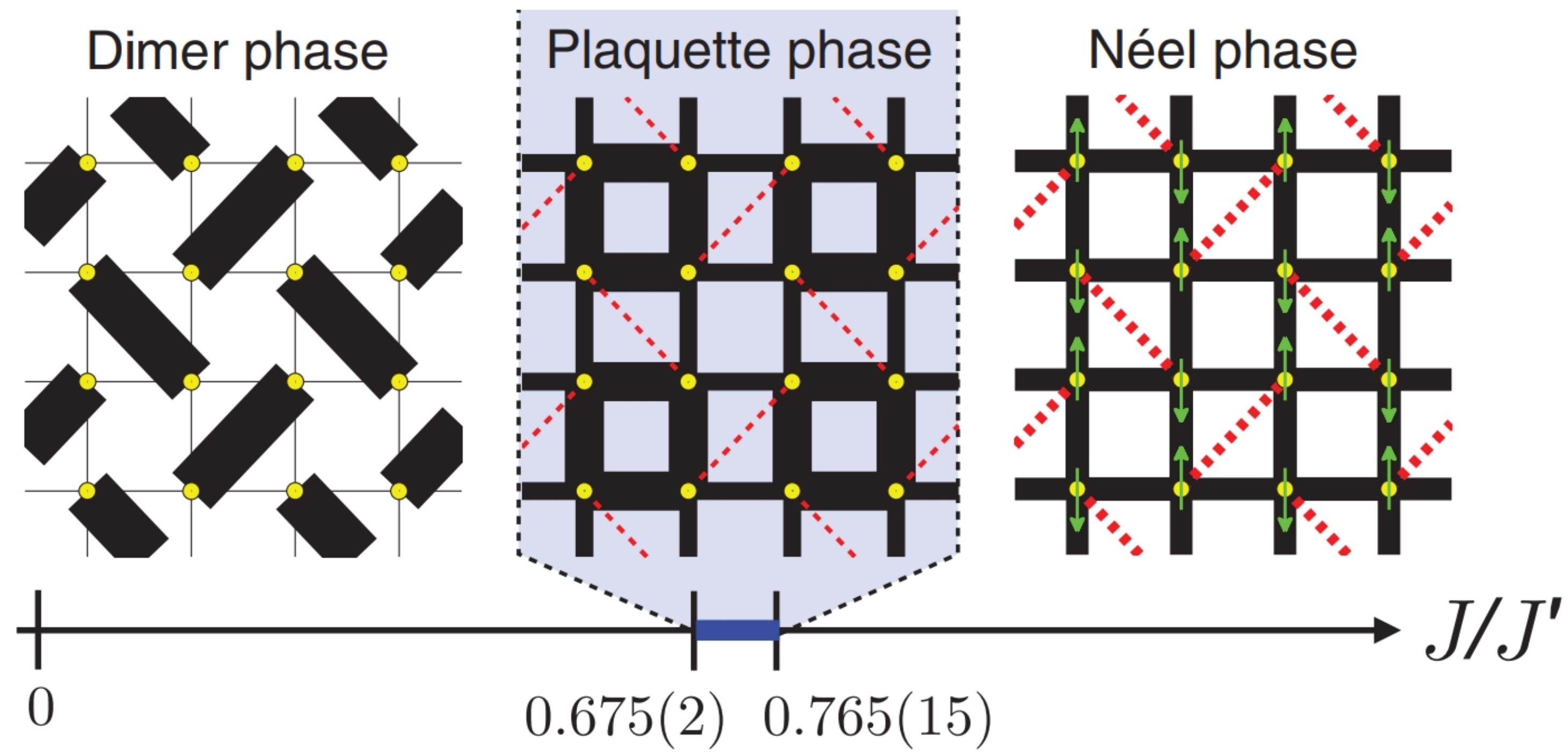}
\caption{\label{ssmpd} Zero temperature phase diagram in the SSM as a function of the ratio of the inter-dimer to the intra-dimer coupling $J/J'$. Adapted from Ref. ~\cite{Corboz_PRB_2013}.
}
\end{figure}

The Shastry-Sutherland model (SSM), proposed by Shastry and Sutherland in 1981 as a toy model~\cite{Shastry_1981},
has attracted significant attention for its intriguing physical properties.
The structure of the model is illustrated in Fig.~\ref{ssm}, which
features intradimer AFM interactions $J'$ where adjacent dimers align perpendicularly,
and the interdimer AFM interactions $J$ among the neighboring sites, in a square lattice.
Due to competing $J$ and $J'$, the model contains an exactly solvable dimer singlet (DS) ground state
in the large $J'$ limit, and an AFM ordering in the large $J$ limit by the first glance.
With improved numerical simulations, as the ratio $\alpha = J/J'$ increases, the model exhibits three distinct ground states, as shown in Fig.~\ref{ssmpd}.
For $0 \leq \alpha \leq 0.675$, the ground state corresponds to a DS phase~\cite{Shastry_1981, Koga_PRL_2000}.
For large $\alpha$, the ground state corresponds to an AFM state~\cite{Manousakis_RMP_1991}. For $0.675 \leq \alpha \leq 0.765$, the ground state
orders in a VBS state, namely a plaquette singlet (PS) state~\cite{Koga_PRL_2000, Corboz_PRB_2013}
where four spins within a plaquette form a local singlet state.

Theoretical studies suggest that a DQCP may emerge in the QPT between the VBS phase and the AFM phase at $\alpha{\approx}0.765$,
accompanied by a change in the Berry phase~\cite{Mcclarty_NP_2017}.
Both the PS and the AFM phases are spontaneous symmetry breaking state: the PS phase breaks $Z_2$ lattice translational symmetry, while the AFM phase breaks spin SU(2) symmetry.
If the PS--AFM phase transition is continuous, fractionalized excitations associated with deconfinement may exist, and an emergent, enhanced symmetry
could arise. However, the nature of this phase transition remains contentious. Some studies propose a second-order transition~\cite{Koga_PRL_2000,Lee_PRX_2019}, while others suggest a weakly first-order transition~\cite{Corboz_PRB_2013,Xi_PRB_2023}. A gapless QSL could also exist between the PS and AFM phase~\cite{Yang_PRB_2022, wang_CPL_2022}.
Note that the DS--PS transition at $\alpha{\approx}0.675$ could not be a DQCP, as the DS phase is not a spontaneous symmetry breaking state.

\subsection{$J_1-J_2$ model on the honeycomb lattice}

\begin{figure}[t]
\centering
\includegraphics[width=6cm]{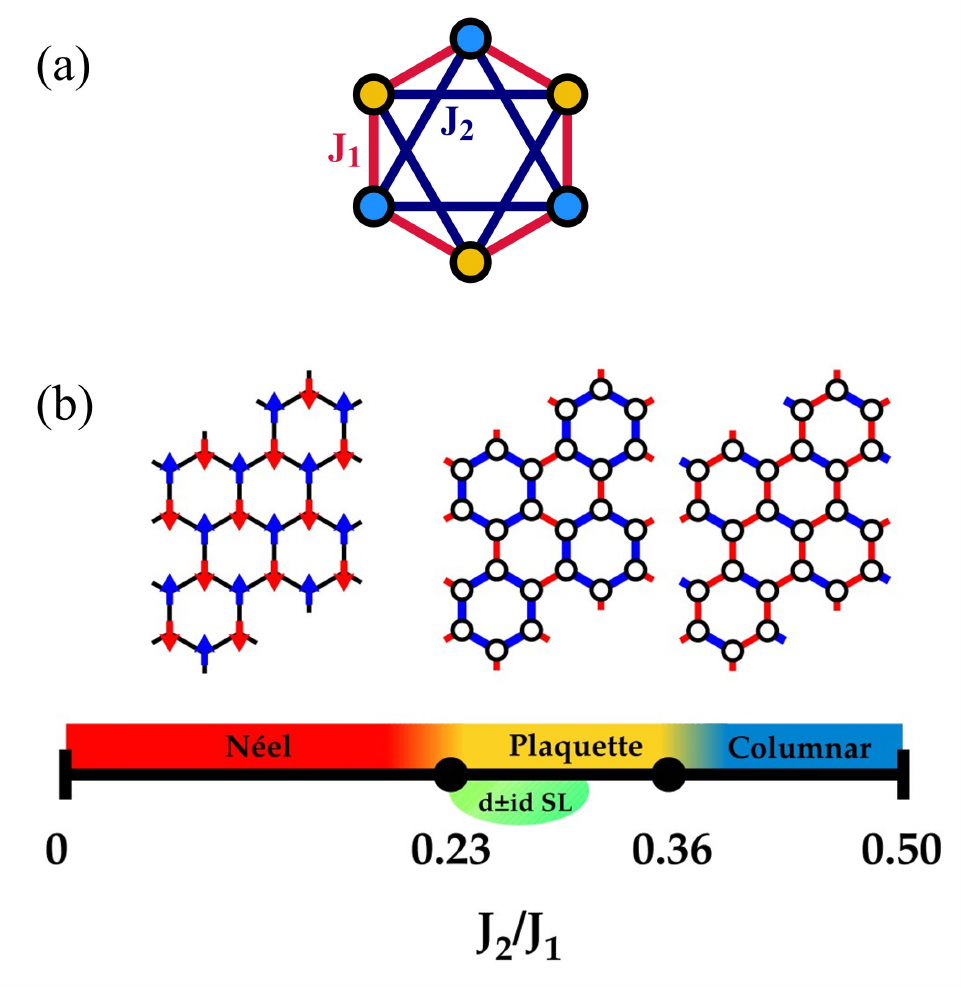}
\caption{\label{honeycomb}
(a) Schematic illustration of the interactions in the $J_1-J_2$ Heisenberg model
on the honeycomb lattice.
(b) Phase diagram of the spin-$1/2$, $J_1-J_2$ model.
The ground state exhibits N{\'e}el, plaquette and columnar dimer orders, with critical points at $J_2/J_1$=0.23 and 0.36.
Sites connected by blue lines form local singlet.
A region of $d\pm id$ QSL with competitive energy is marked by a green ellipse. Adapted from Ref.~\cite{Ferrari_PRB_2017}.
}
\end{figure}

The spin-1/2 $J_1-J_2$ Heisenberg model on the honeycomb lattice is defined by
\begin{equation}
{\mathcal H} = J_1\sum_{\langle ij \rangle}\mathbf{S}_i \cdot \mathbf{S}_j+J_2\sum_{\langle\langle ij \rangle\rangle}\mathbf{S}_i \cdot \mathbf{S}_j ,
\label{hc}
\end{equation}
where $\langle ij \rangle$ and $\langle\langle ij \rangle\rangle$ represent nearest neighboring and next-nearest neighboring bonds, respectively, as shown in Fig.~\ref{honeycomb}(a).
This model has been extensively investigated using techniques such as density matrix renormalization group (DMRG) calculations, coupled-cluster methods, and Monte Carlo simulations~\cite{Farnell_PRB_2011, Ganesh_PRL_2013, Bishop_JP_2013, Ganesh_PRB_2013, Pujari_PRL_2013, Ferrari_PRB_2017}.

The phase diagram of the model, shown in Fig.~\ref{honeycomb}(b) exhibits a rich interplay of quantum phases as the $J_2/J_1$ ratio increases.
For $J_2/J_1 < 0.23$, the ground state exhibits conventional N{\'e}el order. As $J_2/J_1$ increases beyond 0.23, the system transitions
into a plaquette VBS phase, where the six spins in a closed hexagonal unit form a singlet state, resulting in broken translational symmetry.
Interestingly, a gapless $Z_2$ spin liquid, known as the $d \pm id$ state, has a competitive variational energy
in this regime and may be stabilized by longer-range interactions or ring-exchange terms~\cite{Ferrari_PRB_2017}.
For $J_2/J_1 > 0.36$, the ground state become a columnar dimer VBS.
The transition between the N{\'e}el order and the plaquette VBS phase is continuous and may
correspond to a DQCP. Moreover, DMRG simulations suggest that the transition between the plaquette
and columnar dimer states is also continuous, potentially representing a second DQCP in this system~\cite{Ganesh_PRL_2013}.

Experimental investigations into spin-$1/2$ honeycomb lattice compounds have started to shed light on these theoretical predictions.
For example, Wessler {\it et al.} studied YbBr$_3$ using neutron scattering~\cite{Wessler_npj_2020}.
Their findings highlighted competition between nearest-neighbor and next-nearest-neighbor exchange interactions, with continuum
excitations attributed to localized plaquette excitations. While the precise value of $J_2$ remains undetermined,
these results provide compelling evidence for a DQCP in a frustrated honeycomb lattice.
Other spin-$1/2$ honeycomb lattice compounds with Heisenberg interactions, such as $\beta$-Cu$_2$V$_2$O$_7$~\cite{Tsirlin_PRB_2010} and InCu$_{2/3}$V$_{1/3}$O$_3$~\cite{Iakovleva_PRB_2019}, also offer opportunities for studying DQCPs. In these systems, external controls such as pressure, magnetic fields, or chemical substitutions may allow for precise tuning of the $J_2/J_1$ ratio, potentially facilitating the experimental realization and characterization of DQCP phenomena.

\subsection{Hubbard model on the kagome lattice}

\begin{figure}[t]
\centering
\includegraphics[width=8.5cm]{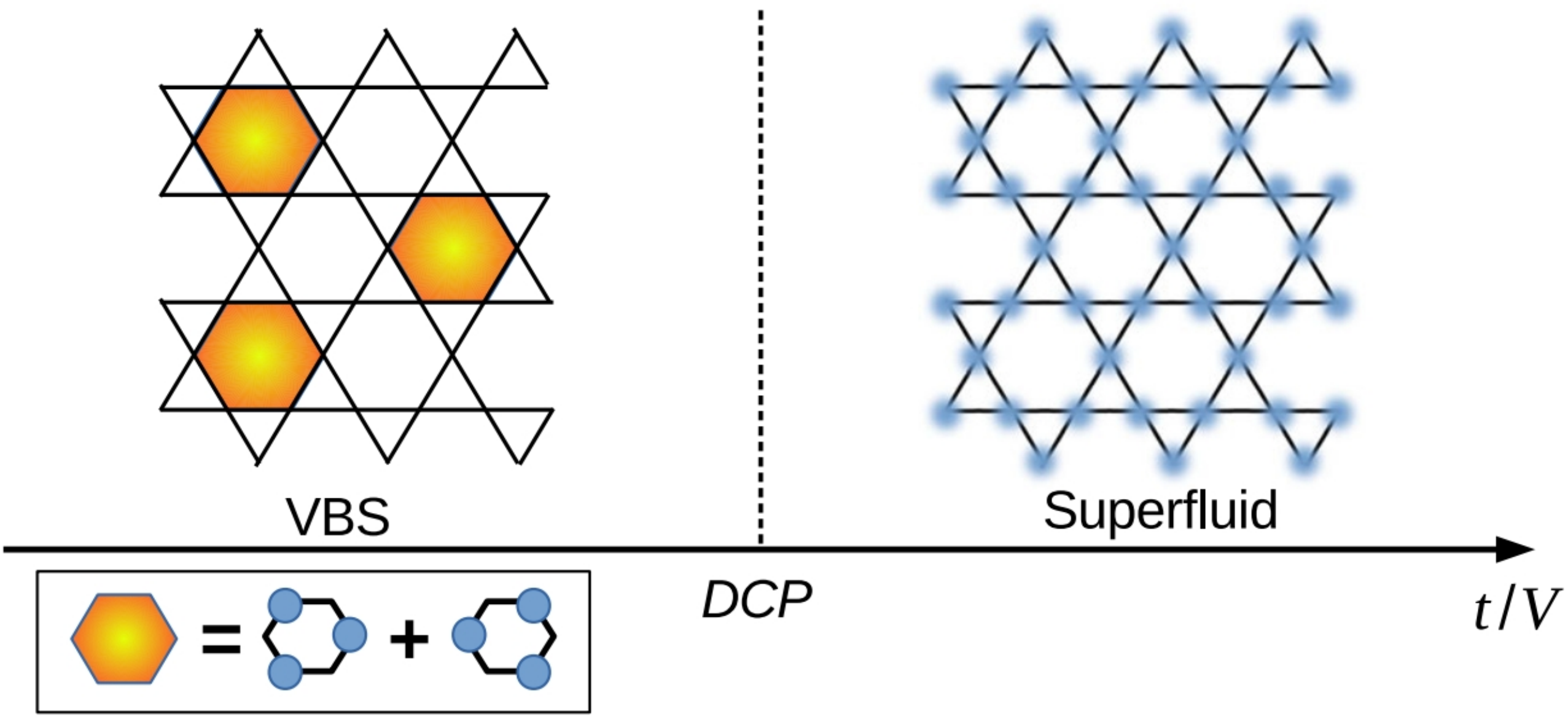}
\caption{\label{kagome}
Phase diagram of the extended Hubbard model of hard core
bosons on the kagome lattice described by Eq.~\ref{ekagome} at 1/3 filling. A DQCP separates a VBS and a superfluid phase. Adapted from Ref.~\cite{Zhang_PRL_2018}.
}
\end{figure}

The kagome lattice also provides a fertile platform for studying DQCPs~\cite{Zhang_PRL_2018,ZhangXF_PRB_2024}. A widely studied
example involves the extended hard-core Bose-Hubbard model, which is described by the Hamiltonian:
\begin{eqnarray}
 {\mathcal H} & = -t& \sum_{\langle ij \rangle} (b^+_ib_j+b^+_jb_i)+V\sum_{\langle ij \rangle} n_in_j,
 \label{ekagome}
\end{eqnarray}
where $t$ ($>$0) and $V$ ($>$0) represent the hopping amplitude and nearest-neighbor repulsive interaction, respectively. The operators $b^+_i$ and $b_i$ denote the creation and annihilation of hard-core bosons.

At a filling of $n=1/3$, the interplay between frustration and quantum fluctuations stabilizes distinct quantum phases.
In the strong-coupling regime with $V\gg t$, the kagome lattice stabilizes a VBS phase.
As shown in Fig.~\ref{kagome}, alternating yellow hexagons, each enclosing six spins, form resonating singlets,
while the remaining spins are ferromagnetically aligned with each other, leading to the breaking of translational symmetry.
In contrast, in the weak-coupling regime with $V \ll t$, the system transitions into a superfluid phase, with long-range phase coherence breaking the global U(1) symmetry.

At the critical point, where $t/V \approx 0.1303$, a continuous DQCP may exit to separate theses two phases.
This critical point exhibits several remarkable features: fractionalized excitations (spinons) and emergent U(1) gauge fields dominate the low-energy physics; an anomalous scaling behavior distinguishes the DQCP from conventional transitions; and the system exhibits an enhanced U(1) symmetry.

\subsection{$J-K$ model on one-dimensional spin chains}

DQCPs were originally proposed in 2D systems. However, their precise nature in such systems remains elusive due to both theoretical and numerical challenges
in frustrated systems. Furthermore, the detection of DQCP in 2D typically relies on fine-tuned models, and numerical approaches often struggle to provide
definitive evidence for key features such as fractional excitations, emergent symmetries, and quantum critical scalings.
In contrast, the study of DQCPs in 1D systems has gained attractions in recent years. Techniques such as the Bethe ansatz, low-energy bosonization,
conformal field theory (CFT), and DMRG enable rigorous investigations, and certain models can
be solved exactly~\cite{Huang_PRB_2019, Mudry_PRB_2019, Roberts_PRB_2019, Jiang_PRB_2019, Huang_PRR_2020,  Roberts_PRB_2021, Zheng_PRB_2022, Xi_CPB_2022, Zhang_PRL_2023, Romen_SciPost_2024,WangYN_arxiv_2024}.

\begin{figure}[t]
\centering
\includegraphics[width=7.5cm]{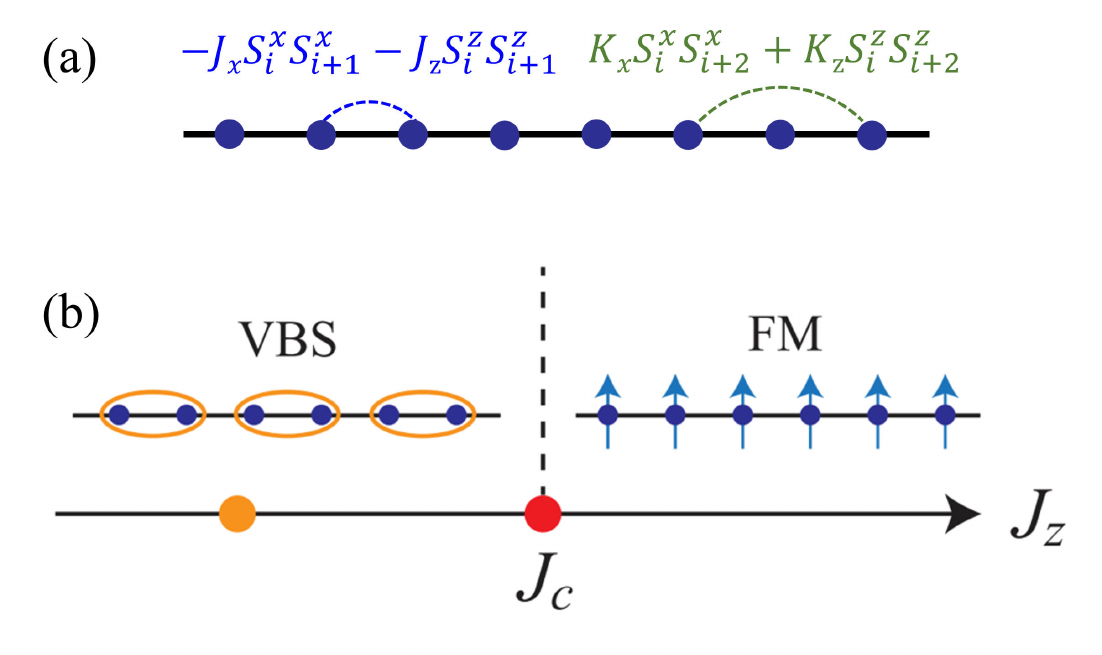}
\caption{\label{1d}
(a) Schematic representation of the spin-$1/2$ chain model described in Eq.~\ref{e1d}.
(b) Phase diagram for the model.
As $J_z$ increases, the ground state transitions from a VBS phase to a FM phase, with $J_c$ marking the critical point. Adapted from Ref.~\cite{Ferrari_PRB_2017}.
}
\end{figure}

Jiang and  Motrunich proposed a 1D spin-$1/2$ model featuring ferromagnetic (FM) nearest-neighbor interactions and AFM second-neighbor interactions~\cite{Jiang_PRB_2019}, depicted in Fig.~\ref{1d}(a).
This 1D model is particularly notable because it allows for an exact solution, providing unambiguous evidence for the DQCP and its associated features. This exact solvability enhances our understanding of the underlying criticality and emergent phenomena. The Hamiltonian is given by
\begin{eqnarray}
 {\mathcal H} & = & \sum_{i} (-J_xS^x_iS^x_{i+1}-J_zS^z_iS^z_{i+1})\nonumber \\
 && + (K_xS^x_iS^x_{i+2}+K_zS^z_iS^z_{i+2}).
 \label{e1d}
\end{eqnarray}
For simplicity, they fixed the second-neighbor AFM interaction $K_x=K_z=1/2$ and set the nearest-neighbor FM interaction $J_x=1$, leaving $J_z$ as the tuning parameter.

The ground state of the system changes with $J_z$, as depicted in Fig.~\ref{1d}(b).
As $J_z$ increases, a transition occurs at $J_c=1.4645$.
For $J_z<J_c$, the system resides in a VBS phase, while for $J_z>J_c$, it enters a FM phase.
These two phases break different symmetries: the VBS phase breaks translation symmetry, while the
FM phase breaks the $Z^x_2$ on-site symmetry. Remarkably, they identified this transition as a
continuous one, characteristic of a DQCP.

\begin{figure}[t]
\centering
\includegraphics[width=5cm, height=5cm]{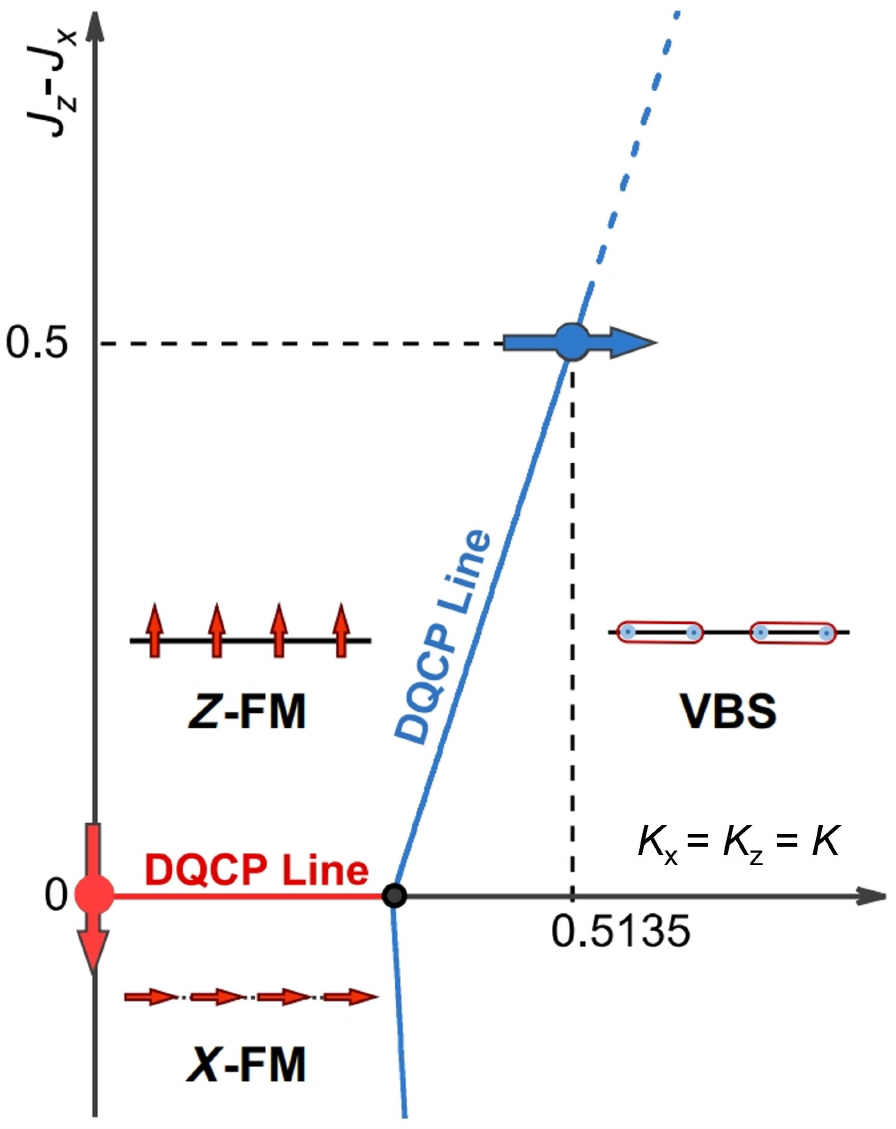}
\caption{\label{1dk} Schematic phase diagram for the spin-$1/2$ chain model described in Eq.~\ref{e1d}. The isotropic line (red line) separates the $X$-FM and $Z$-FM phases, representing DQCP transitions. The blue solid line represents DQCP transitions between the FM and VBS phase. Adapted from Ref.~\cite{Xi_CPB_2022}.
}
\end{figure}

When varying the $K$ parameter~\cite{Xi_CPB_2022}, as shown in Fig.~\ref{1dk}, the ground state for small $K$ values is a FM phase with the moment
aligned along the $Z$ direction when $J_z > J_x$, and along the $X$ direction when $J_z < J_x$. These two FM phases converge at the isotropic line $J_z = J_x$,
where the spin rotational symmetry of the Hamiltonian is enhanced from $Z^x_2 \times Z^z_2$ to a continuous U(1) symmetry. Along this line, the ground state
preserves U(1) symmetry and remains gapless, identifying the transition
as a line of 1D DQCPs. As $K$ increases, the system transitions from the FM phase to the VBS phase. The blue line separating the FM and VBS phases
indicates a line of DQCPs with emergent O(2)$\times$ O(2) symmetry.

Studies have explored DQCPs in other 1D systems, such as the $J_1$-$J_2$
model~\cite{Mudry_PRB_2019}, exactly solvable 1D bosonic model~\cite{ZhangC_PRL_2023},
long-range anistropic Heisenberg model~\cite{Romen_SciPost_2024}, and Rydberg quantum simulators~\cite{Lee_PRL_2023}.
Despite these theoretical progress, no magnetic material has been identified so far to experimentally realize these models.

Here we would like to make comparison of DQCPs in 1D and 2D systems. First,
these two types of DQCPs are described by different theoretical frameworks. The
1D DQCP can be described by a (1+1)D CFT, whereas the 2D DQCP was proposed to be
understood within a (2+1)D CFT. From the theoretical perspective, the (1+1)D CFT
has been well established, while less is known for the (2+1)D one. As a result,
in 2D DQCP there are still many open issues which are
challenging in
both theory and numerical calculations.
Second, DQCPs
are related to the topological nature of excitations. In many 1D chain system,
excitations have topological characteristics, such as kinked excitations in
$S=1/2$ chains and the AKLT state in the $S=1$
Haldane chain. On the other hand, the appearance of Goldstone modes in ordered
phases completely changes the nature of low-energy excitations of a system, and
routes toward topological physics in 2D quantum magnets
remain unclear practically. Nevertheless, the studies of DQCP in 1D system could be illuminating
for 2D systems, for example, by adding magnetic frustration to induce enhanced quantum fluctuations,
which may bring in effective dimensional reduction or topological nature and lead to DQCP resembling
that of the 1D systems.

\section{Experimental investigation of DQCP in the Shastry-Sutherland compound SrCu$_2$(BO$_3$)$_2$}

\begin{figure}[t]
\centering
\includegraphics[width=7cm]{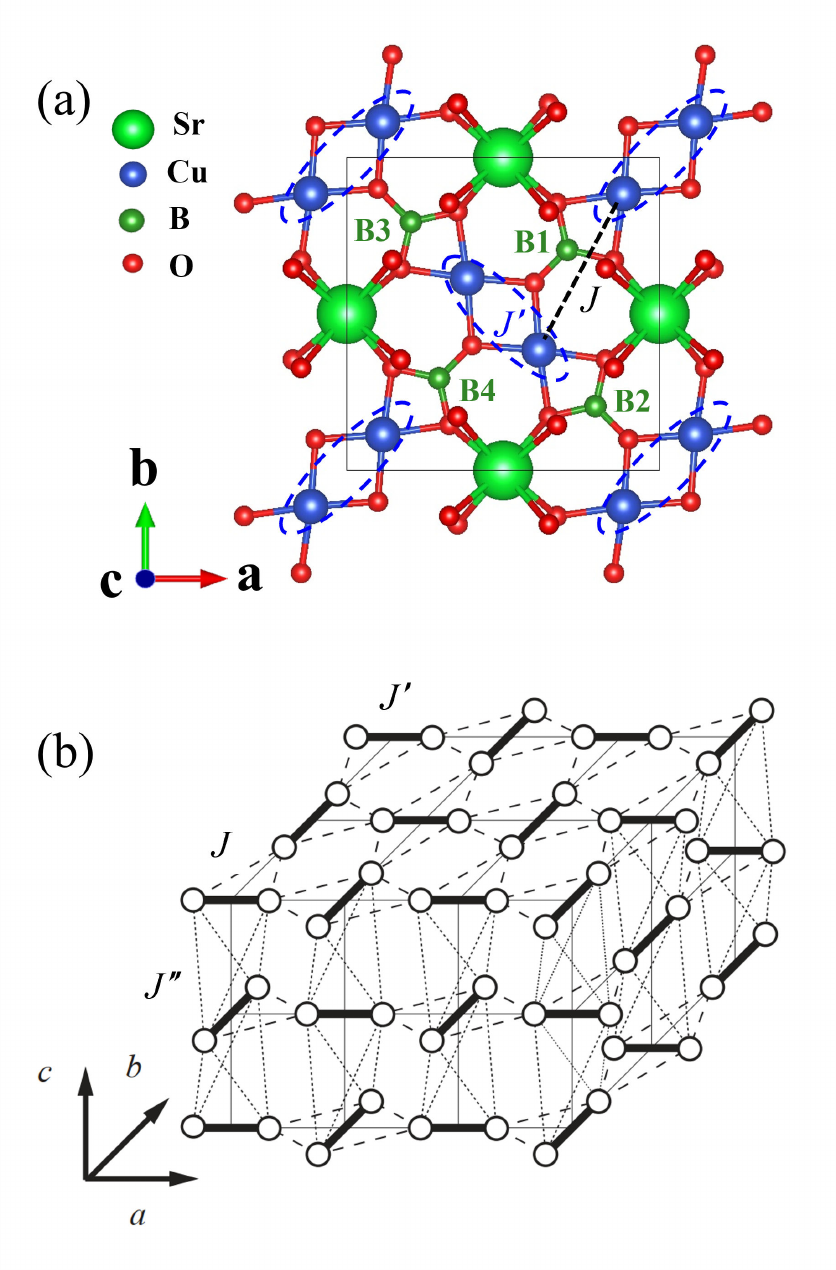}
\caption{\label{scbostruc}
(a) Atomic structure of SrCu$_2$(BO$_3$)$_2$ in the $ab$ plane. Pairs of Cu$^{2+}$
ions form spin dimers (ellipses) with Heisenberg intra-dimer ($J'$) and inter-dimer
($J$) interactions (black dashed lines). Adapted from Ref.~\cite{Cui_Science_2023}. (b) The 3D lattice of SrCu$_2$(BO$_3$)$_2$, adapted from Ref.~\cite{kageyama_PhysB_2003}.
}
\end{figure}

\begin{figure}[t]
\centering
\includegraphics[width=7cm]{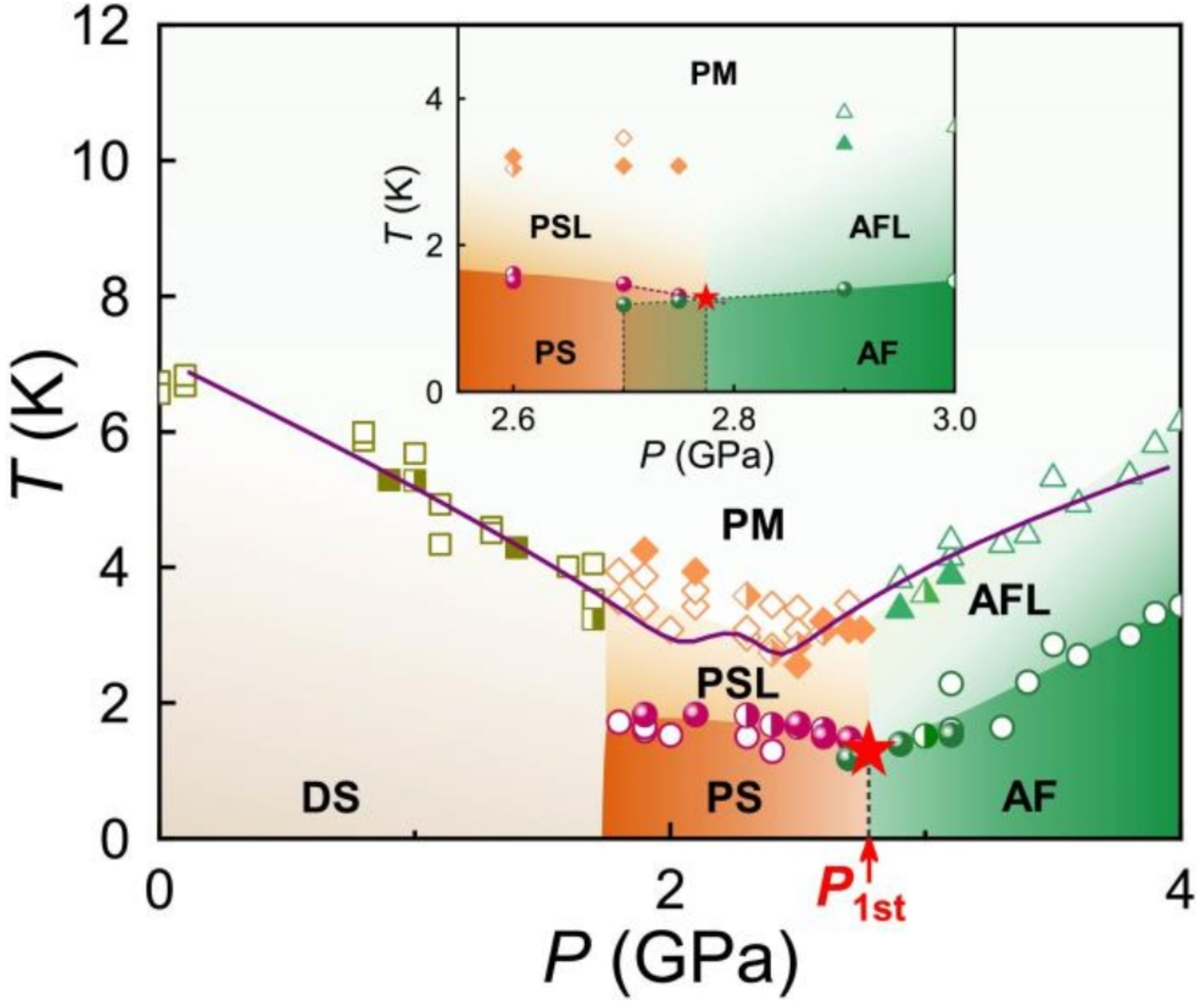}
\caption{\label{cvpd} The ($P$,$T$) phase diagram of SrCu$_2$(BO$_3$)$_2$ obtained from  high-pressure specific heat measurements. The inset shows the transition near the boundary between the PS and AFM phases. The red star marks the critical pressure point of 2.78 GPa, supporting a first-order phase transition, determined by the high onset temperature
of both PS and AFM phases across the transition. Adapted from Ref.~\cite{Guo_2023}.
}
\end{figure}

Research on DQCP has so far been primarily theoretical, with limited experimental progress.
Magnetic materials that enable tuning between distinct, competing symmetry-broken quantum phases usually
require low dimensionality or magnetic frustration.
Pressure is frequently used as an effective method to tune the ratio of microscopic exchange couplings;
however, it usually leads to enhanced 3D magnetic ordering and thus suppresses quantum fluctuations.
Fortunately, a Shastry-Sutherland lattice (SSL) compound SrCu$_2$(BO$_3$)$_2$ offers
a promising platform for investigating such phenomena, due to
its unique bond angle as describe below.

The crystal structure of SrCu$_2$(BO$_3$)$_2$ features spin-1/2 Cu$^{2+}$ magnetic ions
arranged in a square lattice, depicted in Fig.~\ref{scbostruc}(a), representing an ideal realization of the
2D SSM structurally. The intradimer and the interdimer exchange couplings are denoted as $J'$ and $J$.
At ambient pressure, the ground state is a dimer singlet, with
$J'{\approx}$85~K and $\alpha \sim 0.635$,
near the boundary of the PS phase. The intradimer superexchange interaction, governed by the Cu-O-Cu bond angle (97.6$^\circ$),
is highly sensitive to applied pressure~\cite{Sakurai_JPSJ_2018,Boos_PRB_2019}. With applied pressure, the bond angle decreases,
reducing the intradimer AFM interaction within the dimers and driving the system through a series of
QPTs: first from the DS phase to the PS phase, and then to the AFM phase. Lee {\it et al.}~\cite{Lee_PRX_2019} proposed that
the transition between the PS phase and AFM phases in this compound could potentially be a DQCP.
Note that along the $c$-axis, dimers among neighboring layers are connected orthogonally, as shown in Fig.~\ref{scbostruc}(b).
With this configuration, the interlayer coupling is frustrated if the interlayer
coupling is an AFM type~\cite{FoghE_PRL_2024}, and therefore not effective to induce a large $T_N$ under pressure.
Recently, experimentalists have made significant progress in elucidating the
pressure-tuned phase diagram of SrCu$_2$(BO$_3$)$_2$.

\subsection{High-pressure neutron and specific heat measurements on SrCu$_2$(BO$_3$)$_2$}

High-pressure inelastic neutron scattering~\cite{Zayed_NP_2017} and specific heat measurements~\cite{Guo_PRL_2020,Larrea_Nature_2021}
on SrCu$_2$(BO$_3$)$_2$ revealed a gapped phase at pressures above 1.8~GPa, which is consistent with
a PS state.  Further investigations using NMR and neutron scattering spectroscopy
indicate that this ground state corresponds to a PS state. However, it has been found that the PS
state is different from the empty-plaquette (EP) phase as expected from the SSM model~\cite{Zayed_NP_2017,Cui_Science_2023}.
We will return to this point later.

At pressures exceeding 3~GPa, an ordered AFM phase emerges~\cite{Guo_PRL_2020}.
As shown by their established phase diagram in Fig~\ref{cvpd}, a coexistence of PS and AFM phases,
both with high transition temperatures at the transition pressure, was found.
Therefore, the specific heat data supports that this pressure-induced PS--AFM phase
transition is first-order-like~\cite{Guo_2023}.
In general, it is intriguing to determine whether
the QPT is continuous or first-order under high pressure.
Significant challenges are manifold: pressure is usually applied
at room temperature, which does not allow continuous tuning at low temperatures;
pressure hydrostaticity, which is essential to establish a second-order phase
transition, is also difficult to achieve at pressures above 2~GPa.

A recent theoretical proposal has introduced a new method to identify
deconfined quantum criticality.  In a study of the $S=1/2$ AFM SSM, altermagnetism
was observed in the AFM state, characterized by a non-relativistic splitting of two
chiral magnon bands~\cite{ChenHY_arxiv_2024}. Additionally, a Higgs mode was identified
in the longitudinal excitation channel, softening as the system approaches
the AFM-PS transition, indicating nearly deconfined excitations with
a weakly first-order phase transition. This may
reconcile with the experimental observations.

\subsection{Field-induced proximate DQCP investigated by high-pressure NMR}

The magnetic field serves as a complementary and highly controllable tuning parameter for
QPTs in quantum magnets.  Recently, Cui {\it et al.} observed a proximate DQCP
in SrCu$_2$(BO$_3$)$_2$ by NMR study, achieved through the field tuning of
the PS phase under high pressures~\cite{Cui_Science_2023}.

Their high-pressure $^{11}$B NMR study at low fields confirmed the PS phase;
however, with the significant broadening of the NMR satellite spectra,
a full-plaquette (FP) singlet state was suggested, rather than the EP phase.
Subsequent high-field study, with field above 6 T, revealed a field-induced
PS--AFM transition through a Bose-Einstein condensation,
marking a field-induced QPT of this type in this material.
Note that the AFM order is characterized by the onset of the NMR line splits~\cite{Cui_Science_2023}.

\begin{figure}[t]
\centering
\includegraphics[width=8cm]{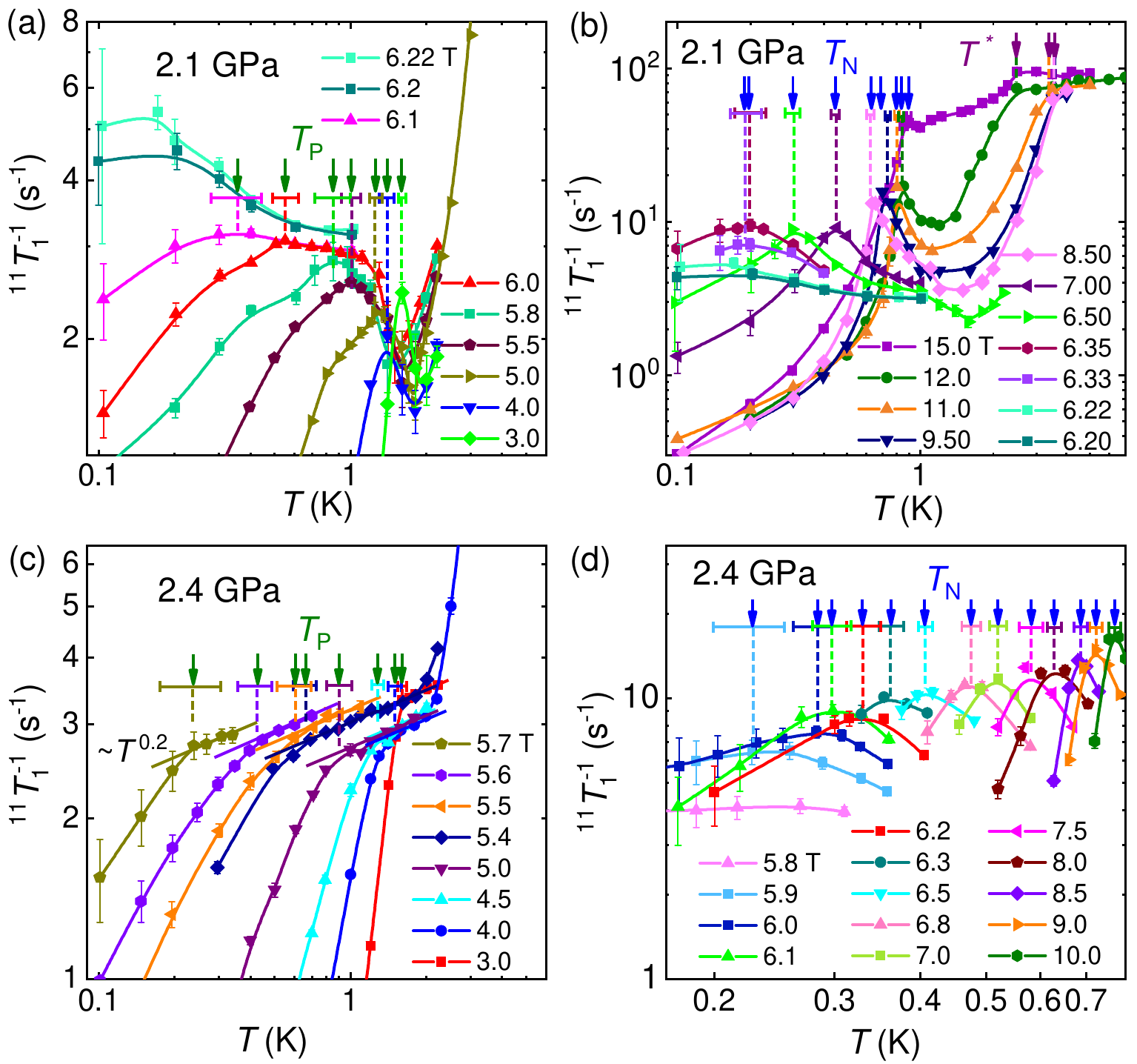}
\caption{\label{t1}
$1/T_1$ of SrCu$_2$(BO$_3$)$_2$ measured at 2.1~GPa and 2.4~GPa. Data at each pressure are separated to show the
PS phase in (a) and (c), and the AFM phase in (b) and (d). Adapted from Ref.~\cite{Cui_Science_2023}.
}
\end{figure}

The spin-lattice relaxation rate $1/T_1$ is a direct probe of low-energy spin fluctuations and offers precise detection
of PS and AFM ordering temperatures at a specific field.
Figures~\ref{t1}(a) and (b) present $1/T_1$ data at $P$ = 2.1~GPa for a range of applied magnetic fields,
grouped into those below and above 6.2 T, corresponding to the low-temperature PS and AFM phases, respectively.
Similarly, Fig.~\ref{t1}(c) and (d) show the $1/T_1$ data at 2.4~GPa, separated at 5.8~T to distinguish
the PS and AFM phases. In the PS phase, the low-temperature $1/T_1$ increases with the field,
whereas in the AFM phase, the low-temperature $1/T_1$ decreases with field.

\begin{figure}[t]
\centering
\includegraphics[width=8cm]{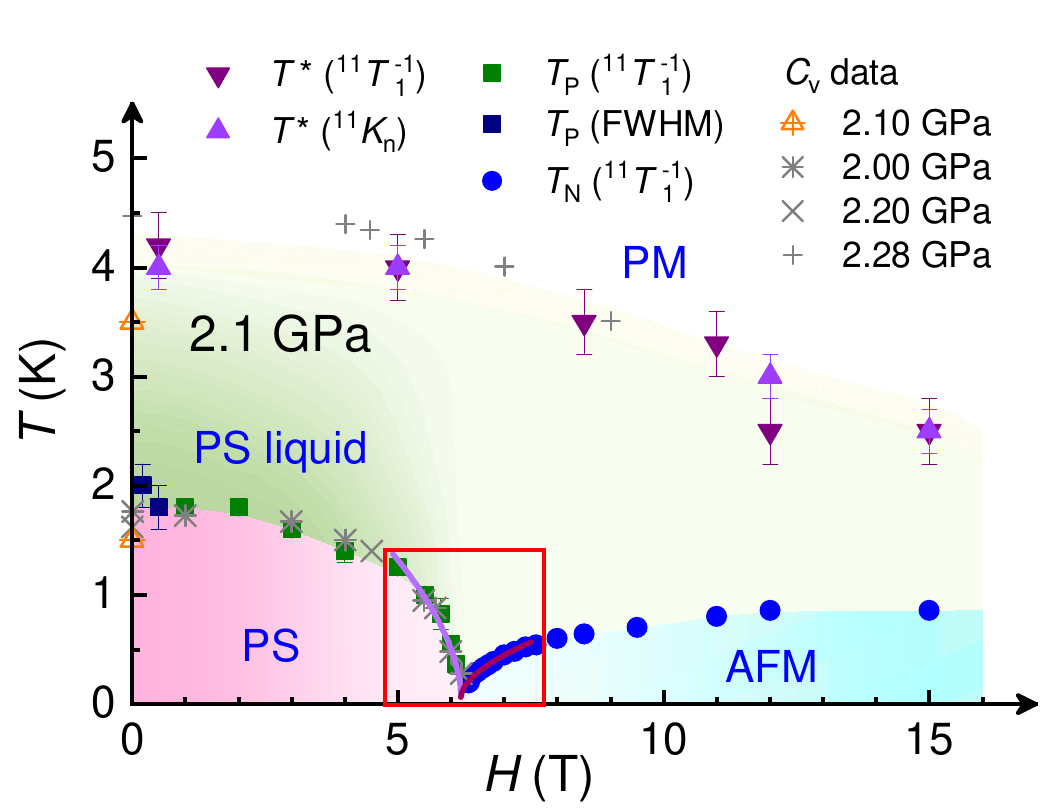}
\caption{\label{scbopd}
Field-temperature phase diagram of SrCu$_2$(BO$_3$)$_2$ at $2.1$~GPa by NMR. The transition temperatures and the
plaquette-liquid crossover temperature $T^*$ are compared with the specific heat data. Adapted from Ref.~\cite{Cui_Science_2023}.
}
\end{figure}

\begin{figure}[t]
\centering
\includegraphics[width=8.5cm]{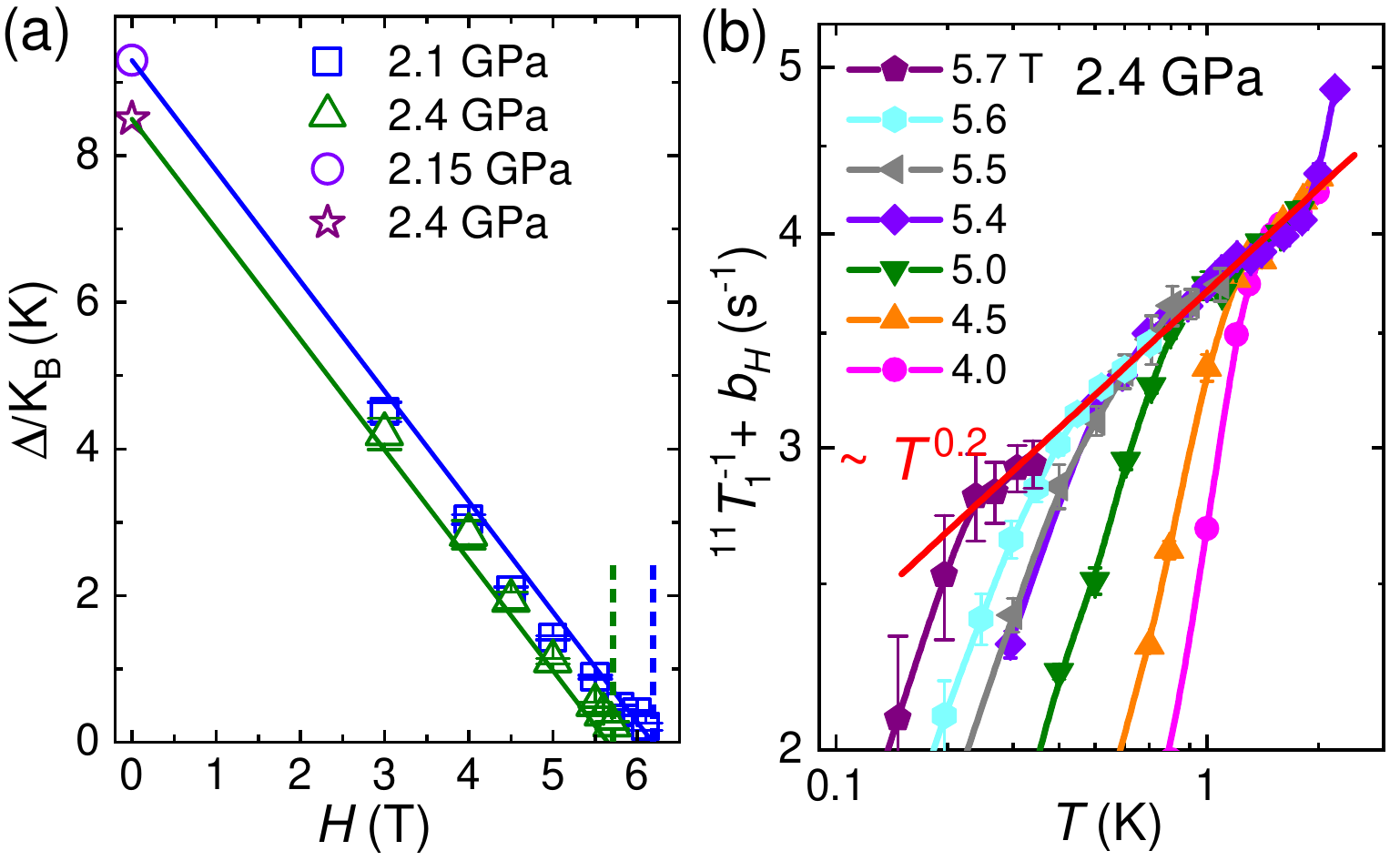}
\caption{\label{gap} Spin gap and critical behavior deduced from $1/T_1$ measurements on SrCu$_2$(BO$_3$)$_2$.
(a) Field dependence of the gap in the PS phase. The lines draw the gap function in the form of $\Delta(H)=\Delta(0)-g\mu_BH$.
(b) Fit of the $1/T_1$ data to the function $1/T_1 = aT^{\eta}-b_H$. Adapted from Ref.~\cite{Cui_Science_2023}.
}
\end{figure}

At $P$~=~2.1~GPa and low fields, Fig.\ref{t1}(a) reveals a broad peak or a sharp kink in $1/T_1$
below 2~K, corresponding to the PS transition temperature $T_{\rm P}$ and the opening of a spin
gap in the PS phase.
At 2.4~GPa, no peak in $1/T_1$ is observed to determine the $T_{\rm P}$ [Fig.~\ref{t1}(c)]; instead,
a sharp crossover from a low-temperature gapped regime to a power-law behavior is found and
used to identify the $T_{\rm P}$.
At higher fields, the sharp peak in $1/T_1$ marks the AFM ordering temperature $T_N$ [see Fig.~\ref{t1}(b) and \ref{t1}(d)].
At 15~T, which is far above the transition field, the $T_N$ is about 0.9~K, which suggests that
the interlayer coupling is indeed very weak in this compound.

With above data, the ($H$, $T$) phase diagram at 2.1~GPa is established as shown in
Fig.~\ref{scbopd}. At zero field, the PS phase transition occurs at 1.8~K,
with the transition temperature gradually decreasing as the magnetic field increases. Beyond 6.2~T, an AFM phase emerges,
and its transition temperature rises with increasing field strength. Near the critical field, the transition temperature
reaches approximately 70~mK, significantly lower than the ordering temperatures of either phase away from the transition point,
suggesting the existence of a proximate DQCP.
The plaquette gap decreases linearly with the magnetic field and vanishes at 6.18~T, as shown in Fig.~\ref{gap}(a),
further supporting the proximity to a continuous phase transition by the field-suppression of the triplet gap.
At 2.4~GPa, the critical field is identified as 5.72~T.

Furthermore, the scaling behavior of the transition temperatures on both sides
exhibits duality with respect to the magnetic field, that is $T_{\rm P,N}{\sim}|H-H_{\rm C}|^{\phi}$
with the same power-law exponent $\phi$ for both the PS phase and the AFM phase.
$\phi$ is obtained as $0.57{\pm}0.02$ at 2.1~GPa and $0.50{\pm}0.04$  at 2.4~GPa.
This is consistent with the onset of an enhanced, emergent symmetry, which makes the
distinction of a conventional QCP and DQCP, given that PS phase breaks the $Z_2$ symmetry
and AFM phase breaks the U(1) symmetry.

With pressure increases from 2.1~GPa to 2.4~GPa, a reduction of the ordered magnetic moment
with pressure is seen at the critical field, which suggests that the
DQCP may be realized at higher pressures. Indeed, at 2.4~GPa,
the $1/T_1$ is already found to follow a quantum critical power-law scaling, $1/T_1 = aT^{\eta}-b_H$.
The scaling exponent $\eta \approx 0.2$ was observed across a temperature window for multiple fields near $H_{\rm c}$ on the
PS side, in contrast to $\eta=0$ expected for conventional QCPs in 2D systems.
Notably,  $\eta$ is consistent with theoretical estimate for an O(4) DQCP~\cite{Qin_PRX_2017}, and slightly lower than
predictions for SO(5) symmetry \cite{Sandvik_PRL_2007,Nahum_PRX_2015}.

\begin{figure}[t]
\centering
\includegraphics[width=8cm]{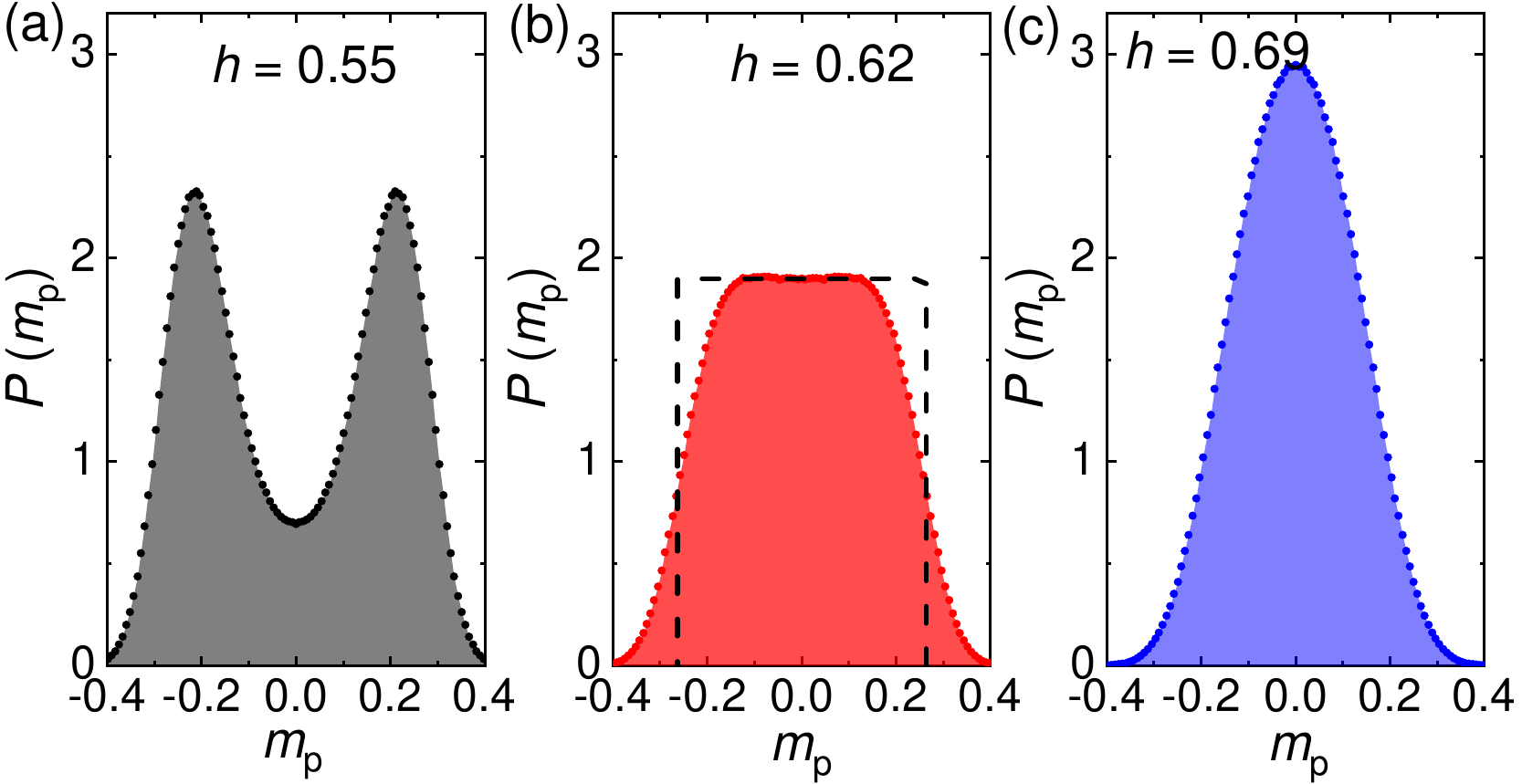}
\caption{\label{jq} Calculated distribution of the plaquette order parameter. Double-peak (a), plateau (b), and single-peak (c) distributions are found, respectively, in the PS phase, at the transition, and in the AFM phase. Adapted from Ref.~\cite{Cui_Science_2023}.
}
\end{figure}

To test a putative O(3) symmetry of the order parameters ($m_x$,$m_y$,$m_p$) in a finite magnetic field, the distribution of the PS order parameter $m_p$ was analyzed
with numerical simulation on a checkerboard JQ (CBJQ) model under magnetic field.
The calculated distribution of the PS order parameter $P(m_p)$ is presented in Fig.~\ref{jq} at three typical fields.
In the PS phase, $P(m_p)$ exhibits non-zero values with a double peak structure as shown in Fig.~\ref{jq}(a),
reflecting the broken $Z_2$ symmetry in the thermodynamic limit. In contrast,
$P(m_p)$ is located at zero in the AFM phase, shown in Fig.~\ref{jq}(c), indicating the absence of PS order.
Interestingly, at the phase transition point,  as shown in Fig.~\ref{jq}(b),
$P(m_p)$ is nearly uniform over a range of $m_p$ values, instead of a three-peak distribution as expected for a conventional
first-order transition with coexisting PS and AFM orders. This uniformity aligns with an O(3) emergent symmetry expected near a DQCP.

\begin{figure}[t]
\centering
\includegraphics[width=6cm]{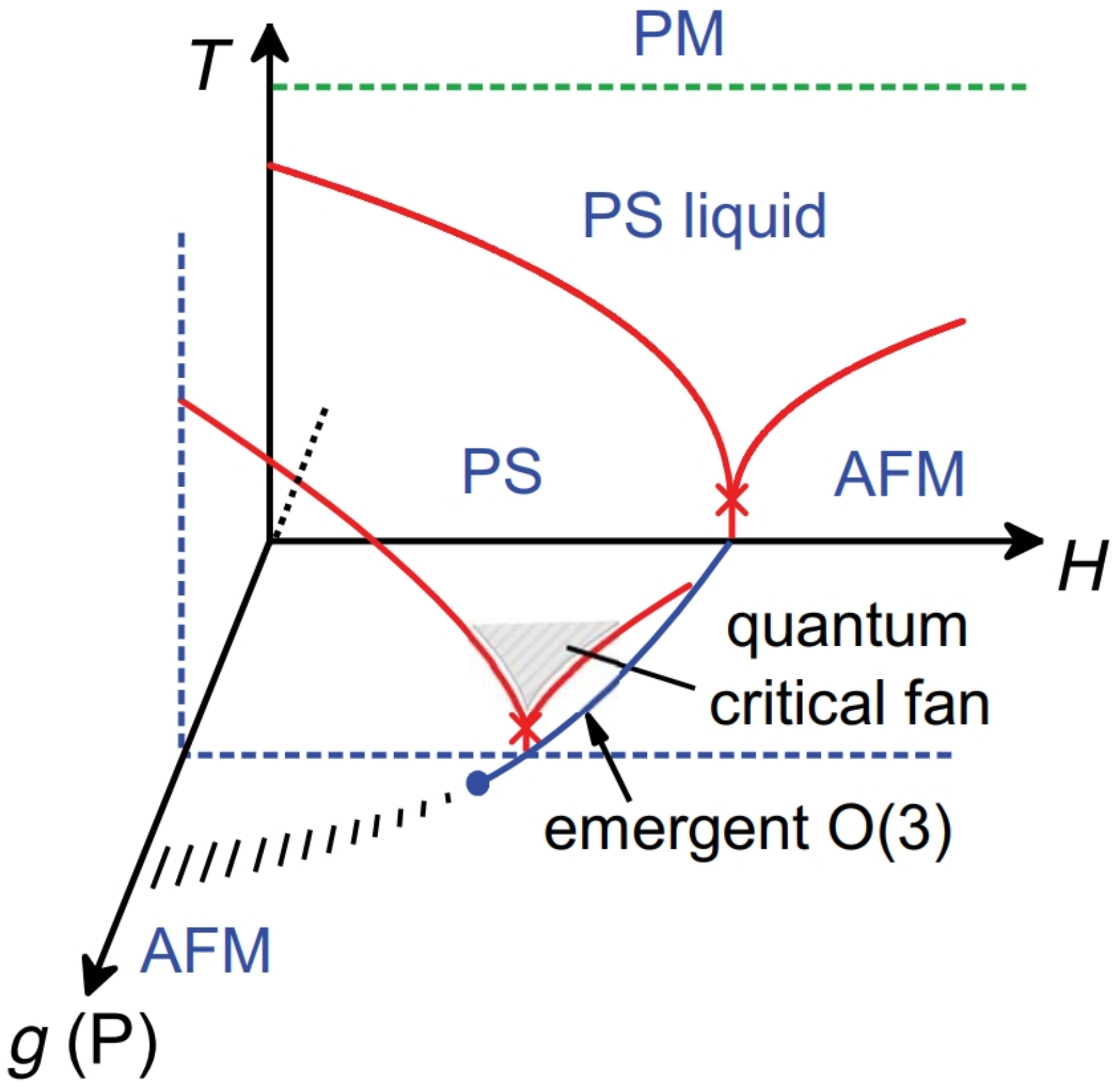}
\caption{\label{pd3d} Schematic phase diagram of SrCu$_2$(BO$_3$)$_2$ in the ($P$,$H$,$T$) parameter space, illustrating key phases such as the PS and AFM phases. The blue dashed lines indicate the case of 2.4~GPa. Adapted from Ref.~\cite{Cui_Science_2023}.
}
\end{figure}

A potential ($P$, $H$, $T$) phase diagram was proposed, as shown in Fig.~\ref{pd3d}, to understand the observation of proximate DQCP in SrCu$_2$(BO$_3$)$_2$  at the above pressures.
As pressure increases, the critical magnetic field for the PS-AFM transition shifts to lower values, and the transition approaches a continuous
QPT, namely, a DQCP.  The dashed line in the phase diagram corresponds to the case of 2.4~GPa, where experimental
results indicate a coexistence of PS and AFM phases below 70~mK. Above 2.4~GPa, the transition becomes continuous, marking the
emergence of a real DQCP. At even higher pressures, additional exotic quantum states, such as  QSL, emerge, further expanding the rich phenomenology of this material.

Note that the approach of DQCP in 2D systems close to the first-order phase transition lines is primarily a practical approach, not a prerequisite.
In 2D systems, according to the Mermin-Wagner theorem, long-range orders can
survive at zero temperature for a system with continuous symmetry. Therefore,
first-order phase transitions among various competing phases are frequently
observed. Close to the first-order phase transition,
adding more tuning parameters may facilitate the realization of DQCPs. For
example, both pressure and magnetic field have been used as tuning parameters to
suppress the first-order phase transition in SrCu$_2$(BO$_3$)$_2$. As a result, searching for DQCPs through first-order transition endpoints
is generally feasible in 2D systems. In contrast, many 1D systems are dominated
by strong quantum fluctuations, without a tendency toward long-range ordering, and
therefore typically favor continuous
transitions described by conformal field theory.

\subsection{Coexisting empty-plaquette and full-plaquette phases}

\begin{figure}[t]
\centering
\includegraphics[width=8.5cm]{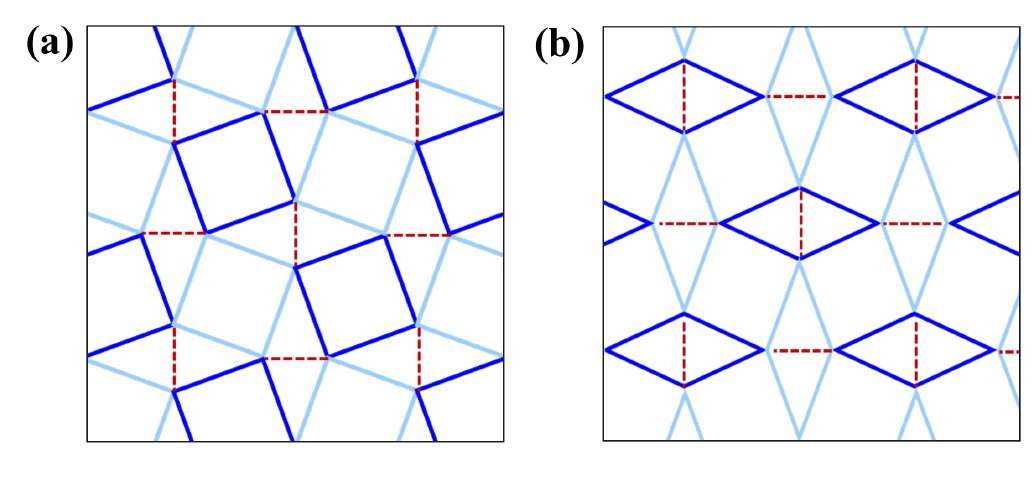}
\caption{\label{epfp} (a) Illustration of the empty-plaquette phase of the SSM, with all dimers (dotted red lines) outside of the local singlets (squares enclosed by thick blue lines).  (b) Illustration of the full-plaquette phase, with half amount of the dimer bonds enclosed in the four-site local singlets (diamonds enclosed by thick blue lines)~\cite{Boos_PRB_2019}.
}
\end{figure}

Note that there is a seeming discrepancy of the PS state for SrCu$_2$(BO$_3$)$_2$ and for the SSM.
In the SSM, an EP configuration, as shown in Fig~\ref{epfp}(a), is expected theoretically~\cite{Boos_PRB_2019,Badrtdinov_PRB_2020,Li_PRL_2023}.
In SrCu$_2$(BO$_3$)$_2$, an FP configuration, as shown in Fig~\ref{epfp}(b), is observed~\cite{Cui_Science_2023}.
These two configurations differ by half amount of the dimers is enclosed by the local singlets in the FP phase,
but none in the EP phase.
Interestingly, recent NMR studies in SrCu$_2$(BO$_3$)$_2$ identified both PS phases coexisting
by a form of phase separation~\cite{Cui_arxiv_2024}. The volume ration of the EP phase increases from
40\% to 70\%, with pressure increasing from 1.9~GPa to 2.65~GPa. Therefore, SrCu$_2$(BO$_3$)$_2$ may be
better described by the SSM at higher pressures.

At 2.4~GPa, the EP phase is also suppressed by an applied magnetic field, leading to the emergence of
the AFM phase. A proximate EP--AFM DQCP is identified at approximately 5.5~T. Near the critical field,
the scaling exponent $\eta=0.6$ was deduced from $1/T_1$, consistent with previous
theoretical predictions for DQCP~\cite{Senthil_PRB_2004}.

The different $\eta$ observed in the FP--AFM and the EP--AFM QPTs may indicate different
universality class of the DQCPs in two cases, which calls for further study.
Earlier SO(5) DQCP models show small values of $\eta$, such as $\eta=0.26$ in
the $J$-$Q$ model ~\cite{Sandvik_PRL_2007} and $\eta=0.33$ in the $J_1$-$J_2$-$J_3$ model~\cite{LiuWY_PRX_2022}.
Recent finite-size tensor network simulations on the SSM have revealed a
continuous PS-AFM transition with $\eta=0.39$, accompanied by emergent O(4) symmetry~\cite{Guzc_PRL_2024}.

\subsection{Other Shastry-Sutherland compounds}

\begin{figure}[t]
\centering
\includegraphics[width=5.5cm]{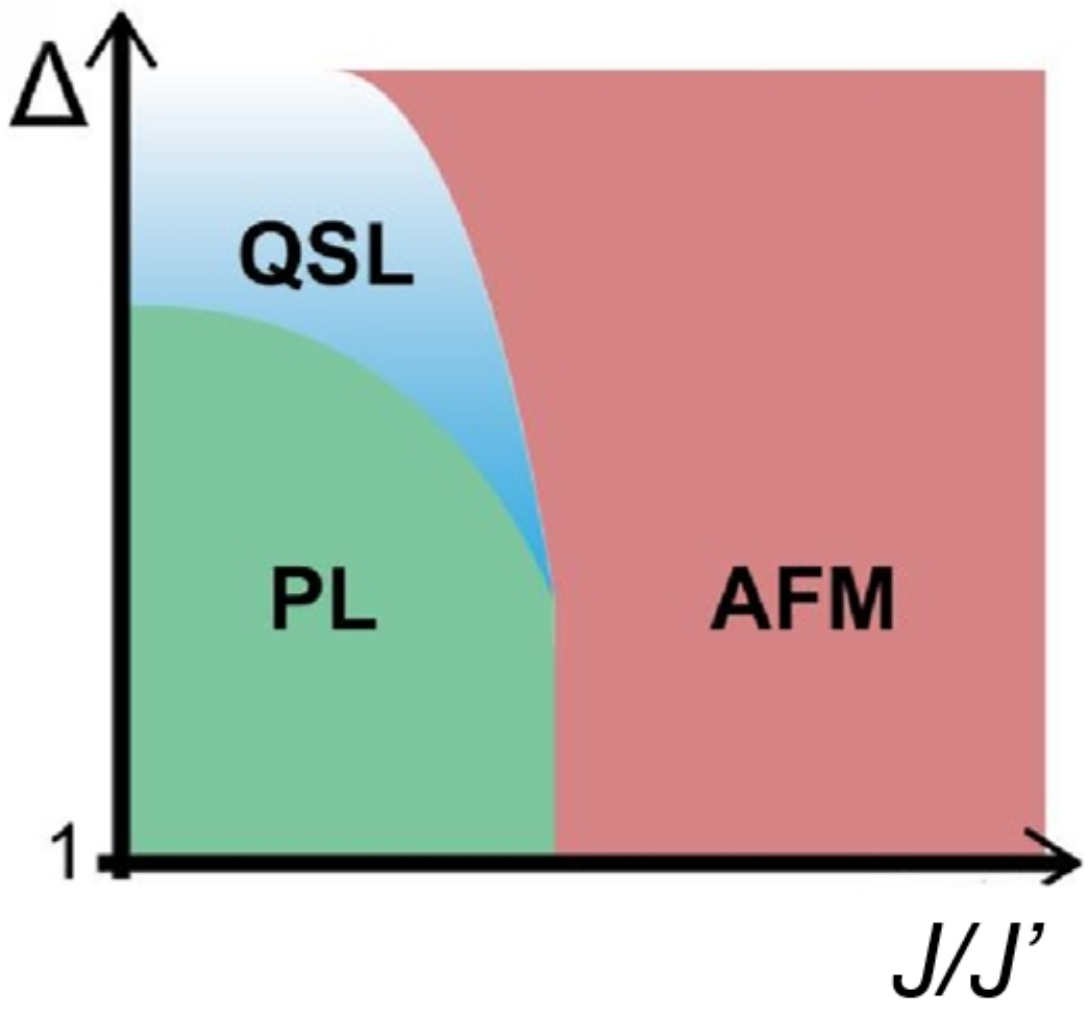}
\caption{\label{xxzpd} Sketched phase diagram of the XXZ SSM.
In the Heisenberg limit ($\Delta = 1$), the transition from the plaquette phase to the AFM phase approaches a DQCP. Increasing spin anisotropy stabilizes a gapless quantum spin liquid (QSL) phase between the plaquette and AFM ground states. Adapted from Ref.~\cite{SunXF_arxiv_2024}.
}
\end{figure}

Other SSL compounds, predominantly containing rare-earth ($R$) ions, have also been identified.
Among binary SSL compounds, $R$B$_4$ stands as a prototypical example~\cite{Blanco_PRB_2006, Okuyama_JMMM_2007, Siemensmeyer_PRL_2008, Ye_PRB_2017},
with research primarily focusing on the field-induced magnetization plateaus.
Among ternary SSL compounds, notable examples include systems of the form $R_2T_2X$, where $T$ denotes
transition metals. Well-studied representatives in this category include Yb$_2$Pt$_2$Pb~\cite{Miiller_PRB_2016},
Yb$_2$Si$_2$Al~\cite{Gannon_PRB_2018}, and U$_2$Pd$_2$In~\cite{Prokes_PRR_2020}.
Furthermore, quaternary SSL compounds,
such as $R$$_2$Be$_2$GeO$_7$ ($R$ = Pr, Nd, Gd-Yb)~\cite{TianZM_inorgchem_2021,Luke_PRB_2024},
$R$$_2$Be$_2$SiO$_7$ ($R$ = Nd, Sm, Gd-Yb)~\cite{ZhouHD_PRM_2024}, Pr$_2$Ga$_2$BeO$_7$~\cite{SunXF_arxiv_2024},
and Ba$R_2T$O$_5$ ($R$ = Pr, Sm, Eu)~\cite{Ishii_JSSC_2020},
are generally insulating, in contrast to the metallic behavior of most binary and ternary SSL compounds.

The insulating nature of these materials offers potential for more effective tuning of both intradimer and
interdimer interactions. However, most of these compounds exhibit nearly equal interaction strengths
within and between dimers, that is, $J/J'{\approx} 1$, leading to the AFM ground state at
ambient pressure. This poses challenges for exploring QPTs with pressure or field tuning.

In the SSM, the QPT between the PS state and the AFM state, associated with a DQCP, occurs in the Heisenberg limit ($\Delta = 1$).
However, many SSL compounds, other than SrCu$_2$(BO$_3$)$_2$,
are closer to the Ising limit, potentially giving rise to different physical phenomena.
Notably, theoretical studies suggest that enhancing the spin anisotropy $\Delta$ stabilizes
a gapless QSL phase between the PS and AFM ground states, as depicted in Fig.~\ref{xxzpd}~\cite{SunXF_arxiv_2024}.

\section{Summary and outlook}

DQCP challenges the conventional Landau-Ginzburg paradigm by describing continuous phase transitions
between two distinct symmetry-broken states. Several theoretical models, especially those with competing
interactions and different lattice geometries, predict the potential existence of DQCPs. While there is
ongoing debate about whether DQCPs exist in these models, exceptions have been noted in
certain 1D systems. The development of new lattice models that inherently contain competing phases
is crucial for further investigating DQCPs and the potential existence of fractional excitations
near QPTs.

In addition to the widely studied AFM-VBS transition, it is important to note that DQCPs may
arise among other phase transitions, such as the 1D FM-VBS transition, different VBS phases, AFM-superconductivity~\cite{Christos_PNAS_2023},
and AFM-charge density wave (CDW)~\cite{LiuZH_PRB_2024},
provided that the phases involved differ not only in their conventional order parameters,
but also in their topological or fractionalized characteristics.

Among potential candidate materials,
SSL compounds, such as SrCu$_2$(BO$_3$)$_2$, have emerged as a promising platform for exploring DQCPs with competing
ordered VBS and AFM states. Notably, SrCu$_2$(BO$_3$)$_2$ has provided clear evidence of a
proximate DQCP via NMR studies, laying a solid foundation for further investigation of
other properties. Nevertheless, direct evidence of spinons through other spectroscopic probes,
such as inelastic neutron scattering, remains challenging due to the stringent requirements for
high field, high pressure, and low-temperature measurements. Furthermore, the search for the
exact existence of DQCPs at higher pressures in this system is also vital, though it may be
hindered by challenges related to hydrostatic pressure conditions.

Additionally, the discovery of new candidate magnetic materials is crucial. Systems with
other competing interactions, such as next-nearest-neighbor couplings, next-next-nearest-neighbor couplings,
or ring-exchange terms, could expand the scope of DQCP research by reaching more competing phases.
Recent studies have also extended DQCP research to non-magnetic systems, including Rydberg quantum
simulators~\cite{Lee_PRL_2023}, trapped ions~\cite{Romen_SciPost_2024}, and ultracold bosons in optical lattices~\cite{Baldelli_PRL_2024}.

While direct experimental observation of DQCPs in  real systems remains challenging,
several other emerging platforms, such as twisted bilayer moir\'{e} superlattice systems composed of
graphene or transition metal dichalcogenides, and iron-based superconductors including BaFe${_2}$(As$_{1-x}$P$_x$)$_2$~\cite{Iye_PRB_2012}
and KFe$_2$As$_2$~\cite{WangPS_PRB_2016} with competing superconductivity and spin-density wave (SDW)/CDW  phases,
offer promising avenues for probing deconfined quantum criticality.

Finally, we comment that direct experimental probe of DQCP is also a challenge.
DQCP exhibits fractional excitations, emergent gauge fields, and enhanced
symmetries.
Spinon excitations lead to continuum spectra for spectroscopic probes such as
inelastic neutron scattering (INS). In quasi-1D materials, spinon excitations have been
confirmed~\cite{Nagler_PRL_1993}, taking advantage of exact numerical
simulations to compare with observed continuum excitations. In 2D systems,
however, the
precise description of the spectral features of the continuum remains
a theoretical and numerical challenge. Alternatively, a combination of different
experimental probes, including INS, NMR, heat conductivity, and others, is
necessary to search for different features of spinons. Second, DQCPs also require
tuning, which usually reduce the accessibility and signal-to-noise ratio of probes.
For example, a combined pressure, high field, and ultra-low temperature approach is
a challenge for INS. Fortunately,
recent numerical progress based on Tensor-network calculations may be helpful to find
fingerprints of fractional excitations in the spectra and scaling behaviors. Even with this,
emergent gauge fields and enhanced symmetries at the DQCP are usually not directly coupled to
the magnetic probes.
Introducing additional tuning or perturbation may be helpful to
trace these quantities and lead to observable changes.

Nevertheless, studying DQCPs will enhance our understanding of emergent phenomena in QPTs, such as
emergent enhanced symmetries, gauge fields, fractional excitations, and the
universality classes of different types of DQCPs. This could also provide valuable
insights for condensed matter physics in general, particularly in strongly correlated
electron systems, and have far-reaching implications for fields
like high-temperature superconductivity, quantum spin liquids, and topological physics~\cite{Mcclarty_NP_2017}.

{\bf Acknowledgments.}---We thank helpful discussions with Prof. Bruce Normand,
Prof. Zhiyuan Xie, and Prof. Zhengxin Liu. This work is supported by
the National Natural Science Foundation of China (Grant No.~12134020 and No.~12374156) and the
National Key Research and Development Program of China (Grant No. 2023YFA1406500).


%

\end{document}